\documentclass[aps,pra,floatfix,twocolumn,superscriptaddress,showkeys,showpacs,10pt]{revtex4-1}
\usepackage{amssymb, amsmath,graphicx,subfigure}
\usepackage{color}
\usepackage{braket}
\usepackage{hyperref}
\usepackage{times}

\begin{document}

\title{Exact quantum dynamics for two-level systems with time-dependent driving}
\author{Zhi-Cheng He 
}

\author{Yi-Xuan Wu 
}
\affiliation{Key Laboratory of Atomic and Subatomic Structure and Quantum Control (Ministry of Education), and School of Physics, South China Normal University, Guangzhou 510006, China}

\author{Zheng-Yuan Xue 
}\email{zyxue83@163.com}
\affiliation{Key Laboratory of Atomic and Subatomic Structure and Quantum Control (Ministry of Education), and School of Physics, South China Normal University, Guangzhou 510006, China}
\affiliation{Guangdong Provincial Key Laboratory of Quantum Engineering and Quantum Materials,  Guangdong-Hong Kong Joint Laboratory of Quantum Matter,   and Frontier Research Institute for Physics,\\ South China Normal University, Guangzhou 510006, China}
\affiliation{Hefei National Laboratory,  Hefei 230088, China}

\begin{abstract}
It is well known that the time-dependent Schrr\"{o}dinger equation can only be solved exactly in very rare cases, even for two-level quantum systems. Thus, finding the exact quantum dynamics under a time-dependent Hamiltonian is not only fundamentally important in quantum physics but also facilitates active quantum manipulations for quantum information processing.
In this work, we present a method for generating nearly infinite analytically assisted solutions to the Schr\"{o}dinger equation for a qubit under time-dependent driving. These analytically assisted solutions feature free parameters with only boundary restrictions, making them applicable in a variety of precise quantum manipulations. Due to the general form of the time-dependent Hamiltonian in our approach, it can be readily implemented in various experimental setups involving qubits. Consequently, our scheme offers new solutions to the Schr\"{o}dinger equation, providing an alternative analytical framework for precise control over qubits.

\keywords{Quantum computation, Quantum control, Analytical solution}

\pacs{03.67.Lx; 03.67.Hk}

\end{abstract}

\maketitle

\section{Introduction}
The search for exact solutions to the time-dependent Schr\"{o}dinger equation has garnered significant interest since the beginning of quantum mechanics. Today, it continues to play a crucial role in various quantum tasks that demand precise control. In particular, the exact evolution of a two-level system under external driving has attracted considerable attention, as it enables accurate qubit control in quantum information processing.
Since a time-dependent Hamiltonian is non-commutative with respect to different times, achieving arbitrary target evolution for a qubit poses a significant challenge. Consequently, exact solutions are typically found only for specific cases, such as the Landau-Zener transition \cite{Landau1932, Zener1932}, Rabi problems \cite{Rabi_Problem}, and quantum control using the hyperbolic secant pulse \cite{Rosen1932}, etc.

Currently, the pursuit of exact dynamics or time-dependent unitary evolutions continues to attract significant attention, as it fundamentally addresses the challenges associated with the lack of precise quantum control. This is particularly relevant in both theoretical research and the experimental realization of quantum control and quantum computation. The primary obstacles include the quench effect, leakage associated with control pulses, and crosstalk among qubits \cite{Exp_Swift_Nearby, quench_1, quench_2, crosstalk_1, crosstalk_2}.

Recently, different schemes for exact quantum dynamics have been proposed 
\cite{Single_Shot_PRL_2013, Fast_Holonomic_Sai_L,Exper_holonomic_Sai_L,Shortcuts_PRL_2012,Shortcuts_Review,shortcuts_z_control,Exactly_Solvable_ADP_2018, Geo_curves_Barnes,Analytically_parametrized_PRA,Analytical_results_PRA,Analytically_Messina_2014,Exact_Holonomic_Transformations,High-speed_driving,High-Order_Harmonic,smooth_exterior,variational_approach}. Notably, an exactly solvable model for a two-level quantum system under a single-axis driving field has been proposed \cite{Barnes2012}, enabling the analytical design of certain types of quantum manipulations \cite{Barnes2013, Economou2015, Exper1_of_Economou2015, Exper2_of_Economou2015}. However, analytical solutions for more general cases remain elusive. Besides, previous solutions also have limitations regarding systematic parameters; for instance, the initial conditions of the pulse shape complicate the design of target evolution operators. Furthermore, obtaining exact dynamics becomes challenging when the driving pulse includes a time-dependent phase in its off-diagonal terms. These limitations hinder the extension of the method to large-scale or long-time quantum control.

Here, we present an analytically assisted scheme for a general time-dependent Hamiltonian of a qubit under driving, capable of generating nearly infinite analytical solutions for its arbitrary dynamics. An arbitrary target evolution under this non-commutative Hamiltonian can be achieved as long as the Hamiltonian can be expressed as a functional of a dimensionless auxiliary function. Notably, these solutions can reduce to well-known analytical solutions, such as those for Rabi problems and the Landau-Zener transition, with specific choices of the auxiliary functions.

Furthermore, we demonstrate the application of our solution in two typical problems. First, we achieve exact quantum dynamics with smooth pulses when manipulating a singlet-triplet (ST) qubit in semiconductor quantum dot systems \cite{QD_review2021}, effectively avoiding the discontinuities in pulse shapes encountered in previous schemes. Second, we implement individual control in multi-level quantum systems with nearby transitions \cite{Exp_Swift_Nearby}. Our analytically assisted solution allows for the design of the Rabi frequency without the constraint of pulse area, enabling us to achieve two desired evolutions in both subspaces with a single pulse. Thus, our scheme offers an analytical-based approach for precise quantum control.

The rest of this work is organized as follows. In Sections II.A and II.B, we present our scheme and its theoretical derivation in detail, focusing on two different control directions within the qubit Hamiltonian. Section II.C discusses specific applications and their corresponding analytical dynamics. Subsequently, in Section III, we address two applications, i.e., tackling the quench problem in experiments and mitigating the leakage of control pulses in multi-level systems. Finally, Section IV provides a brief conclusion of this work.

\section{analytical-assisted solutions} 
In this section, we present our scheme for exactly arbitrary dynamics of qubit system under different driving directions.

\subsection{Solutions with the $\sigma_{z}$ control}

It is well known that the dipole-interacting Hamiltonian for a two-level quantum system, under time-dependent driving and after applying the rotating wave approximation, can generally be expressed as  (assuming $\hbar$ = 1)
\begin{eqnarray}
\label{Hamitonian_General}
H_z=\left(
\begin{array}{cc}
\Delta(t)       &\Omega(t) e^{-i\varphi(t)}\\
\Omega(t) e^{i\varphi(t)}      &-\Delta(t)
\end{array}
\right),
\end{eqnarray}
where $\Omega(t)$ and $\varphi(t)$ label the amplitude and phase of the driven pulse, $\Delta(t)$ is the detuning between the frequency of driven pulse and the energy frequency of a qubit. Its time evolution operator, respected to the unitary condition, can be written as
\begin{eqnarray}
\label{General Time Evolution Operator}
U_{0}=\left(
\begin{array}{cc}
u_{11}       &-u^{*}_{21}\\
u_{21}      &u^{*}_{11}
\end{array}
\right),
\end{eqnarray}
where $|u_{11}|^{2}+|u_{21}|^{2}=1$. 

The first thing we need to consider is how to analyze the entire evolution in the form of algebraic equations, as solving matrix equations can be quite challenging. To address this, we first rotate the entire system into a specific representation, which yields the Schrödinger equation for the evolution operator as
\begin{eqnarray}
i\frac{\partial}{\partial t}(S^{\dagger}U_0)=\left(S^{\dagger}H_{z}S+i\frac{\partial S^{\dagger}}{\partial t}S\right) (S^{\dagger}U_0),
\end{eqnarray}
where the transformation operator $S(t)$ is defined as,
\begin{eqnarray}
S(t)=\exp\left[-i\int^{t}_{0}\Delta(t')dt'\sigma_{z}\right].
\end{eqnarray}
Then, we will convert the matrix equations into algebraic equations for the elements of the evolution operator in Eq.(\ref{General Time Evolution Operator}). To begin, we will introduce new definitions of them as
\begin{subequations}
\begin{align}
v_{11}=&\exp{\left(i\int^{t}_{0} \Delta dt'\right)} u_{11}, \\
v_{21}=&\exp{\left(-i\int^{t}_{0} \Delta dt'\right)} u_{21}.
\end{align}
\end{subequations} Then, we  rewrite the matrix form of Schr\"{o}dinger equation as two algebraic equations as 
\begin{equation}
\label{Relation}
\dot{v}_{11} =i\Omega e^{i\alpha}v_{21}, 
\quad  \dot{v}_{21} =i\Omega e^{i\alpha}v_{11},   
\end{equation}
where $\dot{v}_{11}$ and $\dot{v}_{21}$ label the time derivative of $v_{11}$ and $v_{21}$ , and $\alpha(t)$ is defined as $\alpha(t)=2\int^{t}_{0}\Omega(t')dt'-\varphi+\pi$.
Combining two equations in Eq.(\ref{Relation}), once get $(\dot{v}_{11}/v_{11})(\dot{v}_{21}/v_{21})=-\Omega^{2}$. 

Next, we separate this equation into two distinct equations as,
\begin{equation}
\label{solvable}
\dot{v}_{11}/v_{11}=-i\Omega e^{\kappa(t)},\quad 
\dot{v}_{21}/v_{21}=-i\Omega e^{-\kappa(t)},
\end{equation}
where $\kappa(t)$ is an unknown complex parameter introduced to satisfy the combined equation. Then, we obtain a general solution for $v_{11}$ and $v_{21}$ as,
\begin{subequations}
\label{evopra}
\begin{align}
&v_{11}= \exp\left[i\theta_{1} -i\int^{t}_{0}\Omega(t')e^{\kappa(t')}dt'\right],\\
&v_{21}= \exp\left[i\theta_{2} -i\int^{t}_{0}\Omega(t')e^{-\kappa(t')}dt'\right],
\end{align}
\end{subequations}
where $\theta_{1}$ and $\theta_{2}$ are constants. These constant phases arise from the derivation operation and unknown yet. 

Finally, we deal with these important parameters. Substituting Eq.(\ref{evopra}) into Eq.(\ref{Relation}), we can express $\alpha(t)$ in terms of $\kappa(t)$ as,
\begin{eqnarray}
\label{Alpha}
\alpha(t)=-i\kappa(t)+\theta-2\int^{t}_{0}\Omega(t')\sinh{\kappa(t')}dt'
\end{eqnarray}
where $\theta=\theta_{1}-\theta_{2}$.
Since $\Delta$ and $\varphi$ need to be real, $\alpha(t)$, defined in terms of them, also needs to be real. With this consideration, we can separate the complex function $\kappa(t)$ into its real and imaginary parts, labeling them as $\kappa_R(t)$ and $\kappa_I(t)$. This allows us to separate the right side in Eq.(\ref{Alpha}) into its real and imaginary components. Furthermore, considering the restriction that $\alpha(t)$ is real, it ensures the right side of Eq.(\ref{Alpha}) contains no imaginary terms. Thus, we get,
\begin{eqnarray}
\kappa_{R}(t)=-2\int^{t}_{0}\Omega(t')\sin{\kappa_{I}(t')}\cosh{\kappa_{R}(t')}dt',\end{eqnarray}
Furthermore, we can obtain the derivative form of this equation as follows,
\begin{eqnarray}
\label{Relation of Kappa}
\frac{\dot{\kappa}_{R}(t)}{\cosh{\kappa}_{R}(t)}=-2\Omega\sin{\kappa_{I}(t)}.
\end{eqnarray}
Generally, this parametric equation cannot be solved analytically. However, it is not necessary to do so if the goal is to obtain the evolution operator. By defining a dimensionless function $\chi(t)=\int^{t}_{0}\Omega(t')\sin{\kappa_{I}(t)}$, we aim to establish the relationship between the Hamiltonian and its evolution operator by parameterizing them as functions of $\chi(t)$.

To achieve the above goal, we need to parameterize $\kappa(t)$ as the function of $\chi(t)$. Since $\kappa_{I}(t)$ has already parameterized by $\chi(t)$, we only need to deal with $\kappa_{R}(t)$. This parameterize is achieved by treating Eq. (\ref{Relation of Kappa}) as a differential equation for $\kappa_{R}(t)$ and $\kappa_{I}(t)$. And, they are parameterized as
\begin{subequations} 
\begin{align}
&\kappa_{R}(t)=\ln{[-\tan{(\chi+C)}]}\\
&\kappa_{I}(t)=\arcsin{\frac{\dot{\chi}}{\Omega}},
\end{align}
\end{subequations}
where $C$ is a undetermined constant came from the integral operation of Eq.(\ref{Relation of Kappa}), which will be discussed and determined later. Since we have parameterized $\kappa(t)$ as a function of $\chi$, the relationship between the elements of the evolution operator and $\chi$ has already been established. Introducing $\zeta(t)=\chi(t)+C$ for clarity, the elements of the evolution operator, $u_{11}$ and $u_{21}$, can be calculated and expressed as functions of $\zeta(t)$ as
\begin{subequations} 
\begin{align}
&u_{11}=\exp\left\{i[\theta_{1}+\xi_{-}+\frac{1}{2}(\theta-\varphi-\pi)]\right\}\cos{\zeta}\\
&u_{21}= \exp\left\{i[\theta_{2}+\xi_{+}+\frac{1}{2}(\theta+\varphi+\pi)]\right\}\sin{\zeta},
\end{align}
\end{subequations}
where the parameter $\xi_{\pm}(t)=\int^{t}_{0}\Omega \sqrt{1- \dot{\zeta}^{2}/\Omega^2} \csc{(2\zeta)}dt' \pm \frac{1}{2}\arcsin(\dot{\zeta}/\Omega)$. Notice that We don't need to apply the transformation $S(t)$ again, as we have already obtained the original elements $u_{11}$ and $u_{21}$ necessary to construct the evolution operator. In the above derivation, we have used two identities, i.e., $\sinh{[\ln{(-\tan{x})}]}=\cot{2x}$ and $\sinh{[\ln{(-\tan{x})}]}-\exp{[\ln{(-\tan{x})}]}=\csc{2x}$. Then, the evolution operator can be written as
\begin{eqnarray}
\label{Time Evolution Operator}
U_{0}=\left(
\begin{array}{cc}
e^{i\theta_{1}} e^{i\xi_{-}'}\cos{\zeta}       &-e^{-i\theta_{2}} e^{-i\xi_{+}'}\sin{\zeta}\\
e^{i\theta_{2}} e^{i\xi_{+}'}\sin{\zeta}      &e^{-i\theta_{1}} e^{-i \xi_{-}'}\cos{\zeta}
\end{array}
\right),
\end{eqnarray}
where $\xi_{\pm}'(t)=\xi_{\pm}+ [\theta \pm (\varphi+\pi)]/2$. Next, we address the integral constant $C$. Considering an integral constant can be arbitrary, the function $\zeta(t)$ should allow a non-zero value when $t=0$. However, this setting appears to conflict with the initial condition stipulated in Eq.(\ref{Time Evolution Operator}), which requires $\zeta(0)=0$ to ensure that the evolution operator is the identity when $t=0$. To resolve this contradiction, we modify the evolution operator to satisfy both the arbitrary choice of $C$ and the initial condition. It is given by
\begin{eqnarray}
\label{ut}
U(t)=U_{0}(t)\cdot U_{0}^{\dagger}(0).
\end{eqnarray}
Note that the uncertain phases $\theta_{1}$ and $\theta_{2}$ cancel each other out in the final evolution operator $U(t)$. This is reasonable, as the final solution should not contain any unknown parameters.

Then, we will determine the corresponding Hamiltonian parameterized by $\chi(t)$. Recalling the definition of $\alpha(t)$, we could get $\Delta(t)=[\dot{\alpha}(t)+\dot{\varphi}(t)]/2$. Since $\alpha(t)$ has already been parameterized by $\zeta(t)$, $\Delta(t)$ can be rewritten as a parametric function of $\zeta(t)$ as follows,
\begin{eqnarray}
\label{Omega Form}
\Delta(\Omega,t)=\frac{\Ddot{\zeta}-\frac{\dot{\zeta}\dot{\Omega}}{\Omega}}{2\Omega\sqrt{1-\frac{\dot{\zeta}^{2}}{\Omega^{2}}}}-\Omega\sqrt{1-\frac{\dot{\zeta}^{2}}{\Omega^{2}}}\cot{(2\zeta)}+\frac{\dot{\varphi}}{2}.
\end{eqnarray}
Note that $\zeta(t)$ is limited to avoid the divergence of the terms $\sqrt{1- \dot{\zeta}^{2}/\Omega^{2}}$ and $\cot{(2\zeta)}$, leading to the restriction of its value within the range of (0,$\pi$/2), as shown in Fig. 1. Furthermore, this restriction will limit the evolution time to the order of $1/\Omega$.
Note that $\zeta(t)$ serves as an auxiliary dimensionless function, whereas $\Dot{\zeta}(t)$ is connected to physical quantities which have the unit of frequency, such as $\Omega(t)$ and $\Delta(t)$ in Hamiltonian.
As a result, we obtain a relationship that the evolution operator and time-dependent Hamiltonian are both depend on $\zeta(t)$.
Since the form of $\zeta(t)$ function can be designed almost at will, it allows for the generation of arbitrary evolution operators. In other words, we can design arbitrary $\zeta(t)$ functions and then derive the corresponding Hamiltonian and its evolution operator as an analytical-assisted solution.

\begin{figure}[tbp]
  \label{Parameters FIG}
  \centering
  \includegraphics[width=\linewidth]{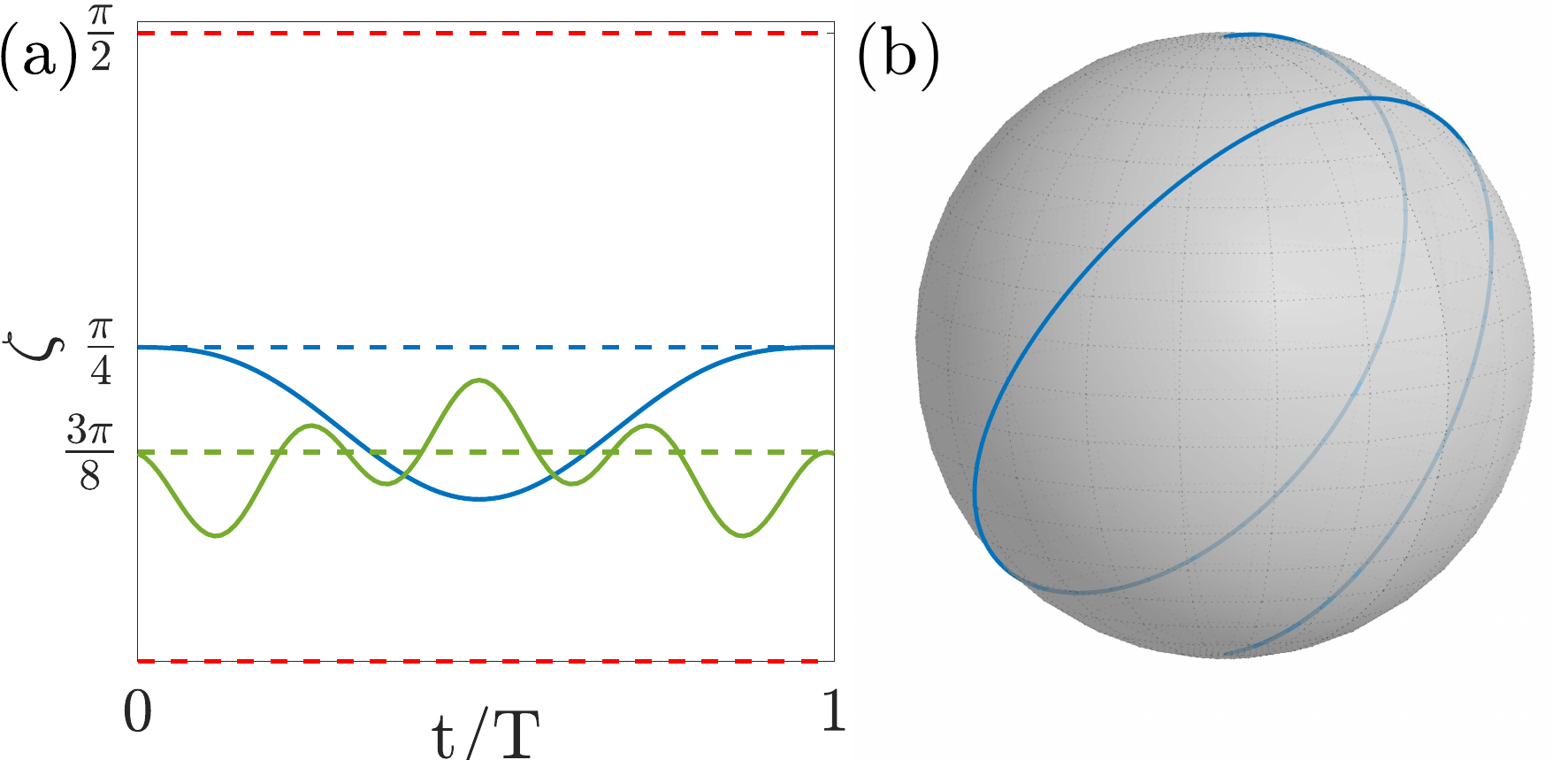}
  \caption{Examples for $\sigma_z$ control. (a) The different types of evolution respect to the  $\zeta(t)$ function. The start and final points of the evolution time was labeled as $t=0$ and $t=T$, respectively. The red dash line describe the limitation of $\zeta(t)$, which is set by the divergence of function $\cot{[2\zeta(t)]}$. The blue solid line describes an example of  population exchange gate.
  The blue dash line with unchanged $\zeta(t)=\pi/4$, which will reduce to Rabi oscillation with constant Rabi frequency. 
  The green solid describes an example of rotations around the axis $\pi/4$ in the x-y plane, and the green dash line with unchanged $\zeta(t)$ implements a rotation around the axis $x+z$ with constant Hamiltonian when $\varphi=0$. (b) The evolution path of the blue solid line in (a), where the population changes from the initial state $\ket{0}$ to the final state $\ket{1}$ in the Bloch sphere, leading to a NOT gate with the parameters being $\Delta=2\pi$ MHz and $T=0.69$ $\mu s$.}
  \label{fig1} 
\end{figure}

\subsection{Transformation for \texorpdfstring{$\sigma_{xy}$}{}  control}

Controlling the transverse $\sigma_{z}$ term typically requires more precise control over the frequency of the driving pulse, which is generally more challenging than controlling the longitudinal $\sigma_{x/y}$ term. Hence, we present the analytical-assisted solution under $\sigma_{x/y}$ control. In this case, to distinguish from Eq. (\ref{Hamitonian_General}), we denote the general Hamiltonian as
\begin{eqnarray}
\label{Hamitonian_Experiment}
H_{xy}=\left(
\begin{array}{cc}
\Delta'(t)       &\Omega'(t) e^{-i\varphi(t)}\\
\Omega'(t) e^{i\varphi(t)}      &-\Delta'(t)
\end{array}
\right).
\end{eqnarray}
This Hamiltonian is indeed equivalent to the Hamiltonian in Eq.(\ref{Hamitonian_General}). The main difference between them lies in the controllable elements. In Eq.(1), we assume control over the $\sigma_{z}$ components to achieve a universal gate set, a scenario commonly found in ST qubit systems \cite{QD_petta2005,Quantum_Dot2004,QD_review2021}. Conversely, the method of controlling over $\sigma_{x/y}$ components  to realize the universality is more prevalent in quantum computation systems  \cite{nv,ions,molecular,superconduct}.
It is important to note that the solution we have obtained thus far only permits control over the  $\sigma_{z}$ components for implementing a universal gate set, while control over the $\sigma_{x/y}$ components remains restricted.
To address this, we introduce a transformation to convert this off-diagonal controllable Hamiltonian into a diagonal controllable one, which is solvable as shown in the previous subsection.
The transformation is
\begin{eqnarray}
U_{R}(t)=\exp{\left[-i\frac{\pi}{4}\cdot
\left(
\begin{array}{cc}
0       &e^{-i(\varphi(t)+\frac{\pi}{2})}\\
 e^{i(\varphi(t)+\frac{\pi}{2})}      &0
\end{array}
\right)\right]},
\end{eqnarray}
then the Hamiltonian in Eq. (\ref{Hamitonian_Experiment}) is changed to
\begin{align}
\label{Hami Trans}
H'&= i\frac{\partial U^{\dagger}_{R}}{\partial t}U_{R}+U^{\dagger}_{R}H_{xy} U_{R}
\nonumber\\
&= \left(
\begin{array}{cc}
\Omega'+\frac{1}{2}\dot{\varphi}    &(-\Delta'+\frac{1}{2}\dot{\varphi})e^{-i\varphi}\\
(-\Delta'+\frac{1}{2}\dot{\varphi})e^{i\varphi}      &-(\Omega'+\frac{1}{2}\dot{\varphi})
\end{array}
\right),    
\end{align}
Now, since $\Omega'(t)$ is arbitrary, comparing with the Hamiltonian in Eq.(\ref{Hamitonian_General}), the evolution operator can be obtained by treating  $(\Omega'(t)+\dot{\varphi}/2)$ as the new parameter of the $\sigma_{z}$ term in the Eq.(\ref{Hamitonian_General}). As a result, the evolution operator can be written as
\begin{eqnarray}
U(t)=U_{R}(t)U'(t)U_{R}^{\dagger}(0).
\end{eqnarray}
where $U'(t)$ is the evolution operator respect to the Hamiltonian in Eq.(\ref{Hami Trans}), which can be obtained according to the $\sigma_{z}$ control case.
And, the controllable off diagonal part in Eq. (\ref{Hamitonian_Experiment}) can be expressed as
\begin{small}
\begin{eqnarray}
\label{Hami Trans Form}
&\Omega'(t)=\frac{\Ddot{\zeta}- \frac{\dot{\zeta}\dot{\Delta}''}{\Delta''}}{2\Delta''\sqrt{1- \frac{\dot{\zeta}^{2}}{\Delta''^{2}}}}-\Delta''\sqrt{1- \frac{\dot{\zeta}^{2}}{\Delta''^{2}}}\cot{(2\zeta)}-\frac{1}{2}\dot{\varphi},
\end{eqnarray}
\end{small}
with $\Delta''=-\Delta'(t)+\frac{1}{2}\dot{\varphi}(t)$.
Furthermore, this analytical quantum dynamics enables the use of phase modulation, in addition to the conventional amplitude shaping, in designing a target quantum control pulse.

\subsection{Results and Dynamics}

As the final result, we write the evolution operator in Eq.(\ref{ut}), 
\begin{eqnarray}
\label{Time Evolution Operator detail}
U(t)=\left(
\begin{array}{cc}
U_{11}       &-U^{*}_{21}\\
U_{21}      &U^{*}_{11}
\end{array}
\right),
\end{eqnarray}
where
\begin{widetext}
\begin{align}
\label{UU Time Evolution Operator}
\nonumber U_{11}= &-e^{\frac{1}{2} i\left[\xi_{-}(0)+2 \xi_{-}(t)-2 \varphi\right]} \cos [\zeta(t)] \cos \left[\zeta(0)\right]-e^{\frac{1}{2} i\left[\xi_{+}(0)-2 \xi_{+}(t)\right]} \sin [\zeta(t)] \sin \left[\zeta(0)\right]\\       \nonumber U_{12}= &e^{-\frac{1}{2} i\left[\xi_{+}(0)+2 \varphi-2 \xi_{-}(t)\right]} \cos \left[\zeta(0)\right] \sin [\zeta(t)]-e^{-\frac{1}{2} i\left[\xi_{-}(0)+2 \xi_{+}(t)\right]} \cos [\zeta(t)] \sin \left[\zeta(0)\right]\\ \nonumber U_{21}= &e^{\frac{1}{2} i\left[\xi_{-}(0)+2 \xi_{+}(t)\right]} \cos \left[\zeta(t)\right] \sin [\zeta(0)]-e^{\frac{1}{2} i\left[\xi_{+}(0)+2\varphi-2 \xi_{-}(t)\right]} \cos [\zeta(0)] \sin \left[\zeta(t)\right]\\      \nonumber U_{22}= &-e^{-\frac{1}{2} i\left[\xi_{-}(0)+2 \xi_{-}(t)-2 \varphi\right]} \cos [\zeta(t)] \cos \left[\zeta(0)\right]-e^{-\frac{1}{2} i\left[\xi_{+}(0)-2 \xi_{+}(t)\right]} \sin [\zeta(t)] \sin \left[\zeta(0)\right]\\
\end{align}
\end{widetext}
with $\xi_{\pm}(t)=\int^{t}_{0}\Omega \sqrt{1- \dot{\zeta}^{2}/\Omega^2} \csc{(2\zeta)}dt' \pm \frac{1}{2}\arcsin(\dot{\zeta}/\Omega)$. Note that, the dynamics of evolution seems depends on various parameters, but, will ultimately depends on one time-dependent term, the auxiliary function $\zeta(t)$. All of other parameters, whatever in the form of Hamiltonian or evolution operator, can be written as the functional of auxiliary function $\zeta(t)$. Such as, the parameter $\xi(t)$, which depend on the integral and the derivative of $\zeta(t)$. Although the
function $\zeta(t)$ itself has no direct physical significance, it serves as a crucial auxiliary function, capable of parameterizing other physical quantities. For instance, its time derivative,
$\dot{\zeta}(t)$, is related to physical quantities with units $T^{-1}$, such as $\Omega(t)$ and $\Delta(t)$ in the Hamiltonian.
Consequently, an almost infinite number of analytical solutions can be derived, as the auxiliary function $\zeta(t)$ can be set nearly arbitrarily, provided it remains within the valid range of $(0,\pi /2)$.

Then, we will demonstrate that analytically-assisted solutions can reduce to well-known analytical solutions of Schd\"{o}inger equation. The dynamics for Landau-Zener transition, such as a ground state is driven through the anti-crossing produced by $\sigma_{x}$ and ultimately returns with probability 1, could be reproduced. In the analytical-assisted solution, considering a ground state at $t=0$, the probability for system to be in ground state at other time can be calculated by Eq.(\ref{ut}), i.e., $|U_{11}(t)|^{2}$. If we choose the auxiliary function $\zeta(t)$ to meet the condition $\dot{\zeta}(0)=\dot{\zeta}(T)=0$, the probability for system in ground state will be $|e^{2i\xi_+}\cos{[\zeta(0)]}\cos{[\zeta(T)]}+\sin{[\zeta(0)]}\sin{[\zeta(T)]}|^{2}$, where T labels the evolution time.
If we further ensure $\zeta(0)=\zeta(T)=0$, a Landau-Zener transition solution can be reproduced. The Hamiltonian also indicate it since $\zeta=0$ will lead to an infinity value of $\sigma_z$ according to Eq.(\ref{Omega Form}), which satisfies the condition of Landau-Zener transition $\sigma_z\gg\sigma_x$. Besides, the Rabi oscillation or other dynamics of time-independent Hamiltonian can also be reproduced if we set the auxiliary function $\zeta(t)$ to be time-independent. Such as, a constant $\zeta=\frac{\pi}{4}$ will lead to a Rabi oscillation while the parameter $\xi(t)$ in Eq.(\ref{Time Evolution Operator}) become the integral of $\Omega(t)$. 

\begin{figure}[tbp]
  \centering
  \includegraphics[width=\linewidth]{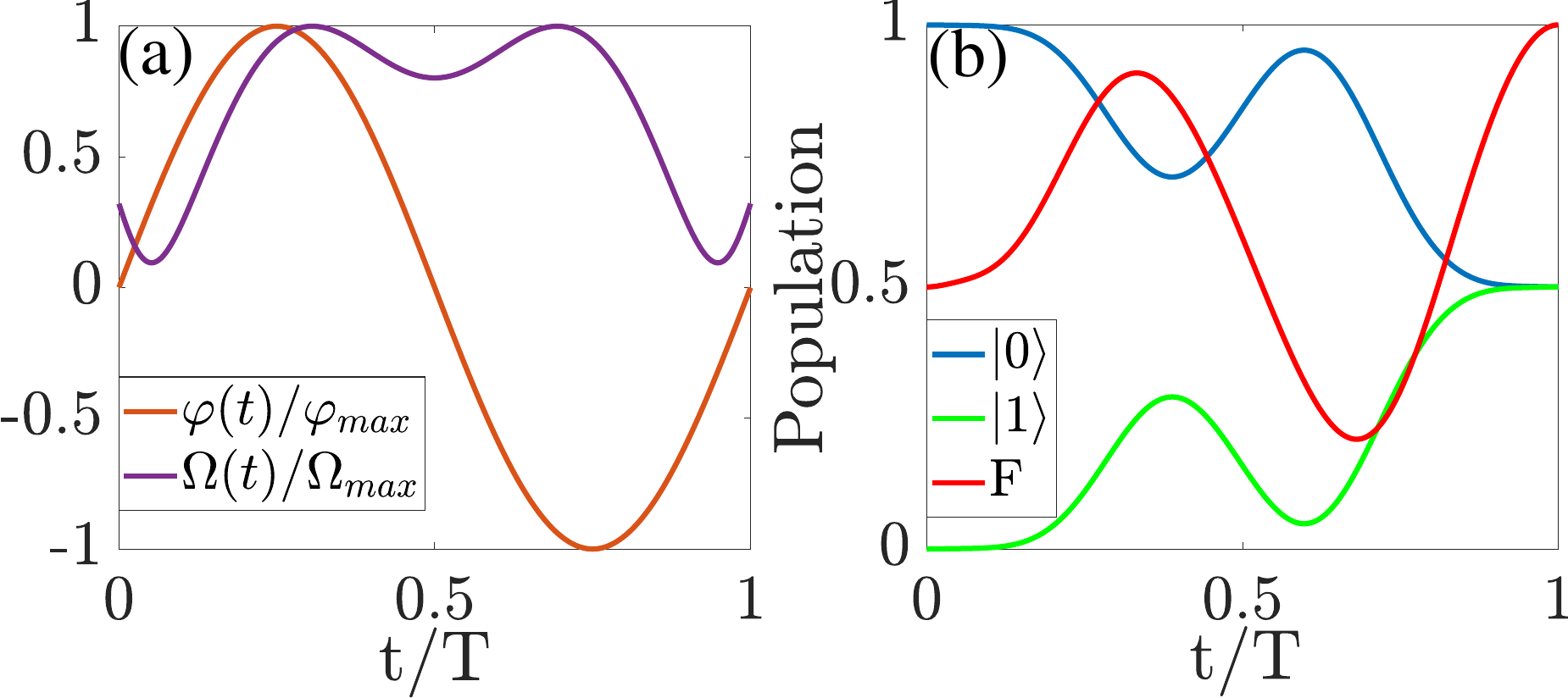}
  \caption{The example of implementing H gate with $\sigma_{x/y}$ control. (a) The time-dependence of the Hamiltonian parameters, including the amplitude $\Omega(t)$ and phase $\varphi(t)$ of the external driving filed. (b) The fidelity and the qubit-state population dynamics for the H gate.}
  \label{fig2}
\end{figure}

Therefore, as a direct application of the solutions, considering a uncontrollable and inevitable time-dependent or time-independent function in the $\sigma_{x/y/z}$ part, achieving a desired two-level evolution can still be accomplished through analytical-assisted solutions. For instance, if the $\sigma_{z}$ part of the Hamiltonian contains an uncontrollable and time-dependent term $\Delta(t)$, it is still possible to derive the desired evolution operator as in Eq. (\ref{ut}) by designing only one function, $\zeta(t)$. Further, the Hamiltonian in other part is also determined since they depend on $\zeta(t)$ and $\Delta(t)$, according to Eq.(\ref{Omega Form}). Moreover, we show more examples with different $\zeta(t)$. It is worth noting that different choices of the parameter $\zeta(t)$ lead to various evolution operators. However, the most common types of evolution operators can typically be obtained under the simpler condition $\zeta(0)=\zeta(T)$, which is easier to determine and calculate other parameters. Thus, we set $\zeta(0)=\zeta(T)$ and express $\zeta(0)$ using a trigonometric series for convenience. Specifically, we set,
\begin{eqnarray}
\label{sin}
\zeta=A_{0}+\sum_{n=1}^{N} A_{n}\sin^{n} \left(a_{n}\pi\frac{t}{T}\right),
\end{eqnarray}
where $A_{n}$ is constant, and the trigonometric series are involved to satisfy $\zeta(0)=\zeta(T)$.
Two examples are shown in Fig.\ref{fig1}, with boundary conditions of $\zeta(t)$ to respect population exchange gates and Hadamard like gates, with parameter sets of $\{A_{0},A_{1},A_{2},A_{3}\}=\{\pi/4, 0, 0, -0.38\}$ and $\{A_{0}, A_{1}, A_{2}, a_{2}, A_{3}, a_{3}\}=\{3\pi/8, 0, -0.22, 4, 0.18, 1\}$, respectively. And the $\zeta(t)$ function is plotted in Fig.\ref{fig1}a. In Fig.\ref{fig1}b, we show an evolution path of the NOT gate by further setting $\Delta=2\pi$ MHz and $T=0.69$ $\mu s$. Similar for $\sigma_{xy}$ control, we design a Hadamard gate when assuming there exists a harmful constant detuning $\Delta'=2\pi$ MHz and the time-dependent phase $\varphi(t)=\sin{(2\pi t/T)}$, as shown in Fig.\ref{fig2}. The Hadamard gate can be obtain when the parameters of $\zeta(t)$ are $\{{A_{0}}, A_{1}, a_{1}\}=\{\pi/8, 0.26, 1\}$ and time $T=0.66$ $\mu s$. The fidelity of the state $\ket{\psi(t)}$ is defined as $F(t)=\lvert \braket{\Psi|\psi(t)} \rvert^{2}$, where $\ket{\Psi}$ labels the ideal target state.

\section{Applications}

In this section, we demonstrate that the analytical solution can be utilized to implement universal gates with smooth pulses, which are well-suited for experimental setups.

\subsection{Gate operation in semiconductor qubits}
The Hamiltonian for ST qubit in semiconductor double quantum dot systems \cite{QD_review2021,Quantum_Dot2004,QD_petta2005} is
\begin{eqnarray}
\label{HST}
H_{ST}(t)=h\sigma_{x}+J(t)\sigma_{z},
\end{eqnarray}
with the qubit basis being defined as $\ket{0}=\ket{T}=(\ket{\uparrow \downarrow}+\ket{\downarrow \uparrow})/\sqrt{2}$ and $\ket{1}=\ket{S}=(\ket{\uparrow \downarrow}-\ket{\downarrow \uparrow})/\sqrt{2}$, where $\ket{\uparrow}/ \ket{\downarrow}$ indicates the spin up/down of an existed electron in left or right dot. $h=g\mu \Delta B_{z}$ and $J(t)$ is the exchange interaction of two dots.
Considering the randomized benchmarking (RB) \cite{Randomized_Benchmarking,RB_Optimal,RB_Process_Tomography} would choose different quantum gates randomly in Clifford gate series to test the quality of gate implementation, these 24 Clifford gate \cite{Clifford_Gates} need to be implemented by the Hamiltonian Eq. (\ref{HST}).. 
In the single-transmon (ST) qubit system, the Hamiltonian comprises only the $\sigma_{x}$ and $\sigma_{z}$ terms, which are used for implementing quantum gates. Typically, the value in the $\sigma_{x}$ term remains time-independent. Previously, to implement quantum gates, it was necessary to keep the time-dependent parameter $J(t)$ constant throughout the evolution, as the dynamics could not be solved if $J(t)$ is time-dependent. Therefore, this manipulation can result in unwanted discontinuities in the pulse profile at the beginning and end of each gate when randomized benchmarking (RB) is performing. These discontinuities can lead to infidelities in the  RB process, as it is challenging to achieve exact realization in experiments. Therefore, we prefer smooth pulse schemes without such breaking points.

We demonstrate that these smooth pulses can be achieved using the analytical-assisted solution. The Hamiltonian in Eq. (\ref{HST}) and Eq. (\ref{Omega Form}) suggests that the time-dependent $sigma_{z}$ control scheme is suitable for the ST qubit system. Since $h$ is time-independent and $\varphi=0$, we obtain the expression of $J(t)$ as a functional of $\zeta(t)$ as
\begin{eqnarray}
\label{J Form}
J(t)=\frac{\Ddot{\zeta}(t)}{2h\sqrt{1-\frac{\dot{\zeta}(t)^{2}}{h^{2}}}}-h\sqrt{1-\frac{\dot{\zeta}(t)^{2}}{h^{2}}}\cot{[2\zeta(t)]}.
\end{eqnarray}
Note that, the evolution operator in Eq.(\ref{Time Evolution Operator}) does not depend on $\Ddot{\zeta}(t)$ directly. Therefore, it is possible to design the suitable $\Ddot{\zeta}(0)$ and $\Ddot{\zeta}(T)$ ensure that the value of $J(t)$ at the start and end points can be a fixed constant (usually zero) in the implementation of all the 24 Clifford gate operations. As the result, a RB process with smooth pulse is obtained. 

\begin{figure}[tbp]
  \centering
  \includegraphics[width=\linewidth]{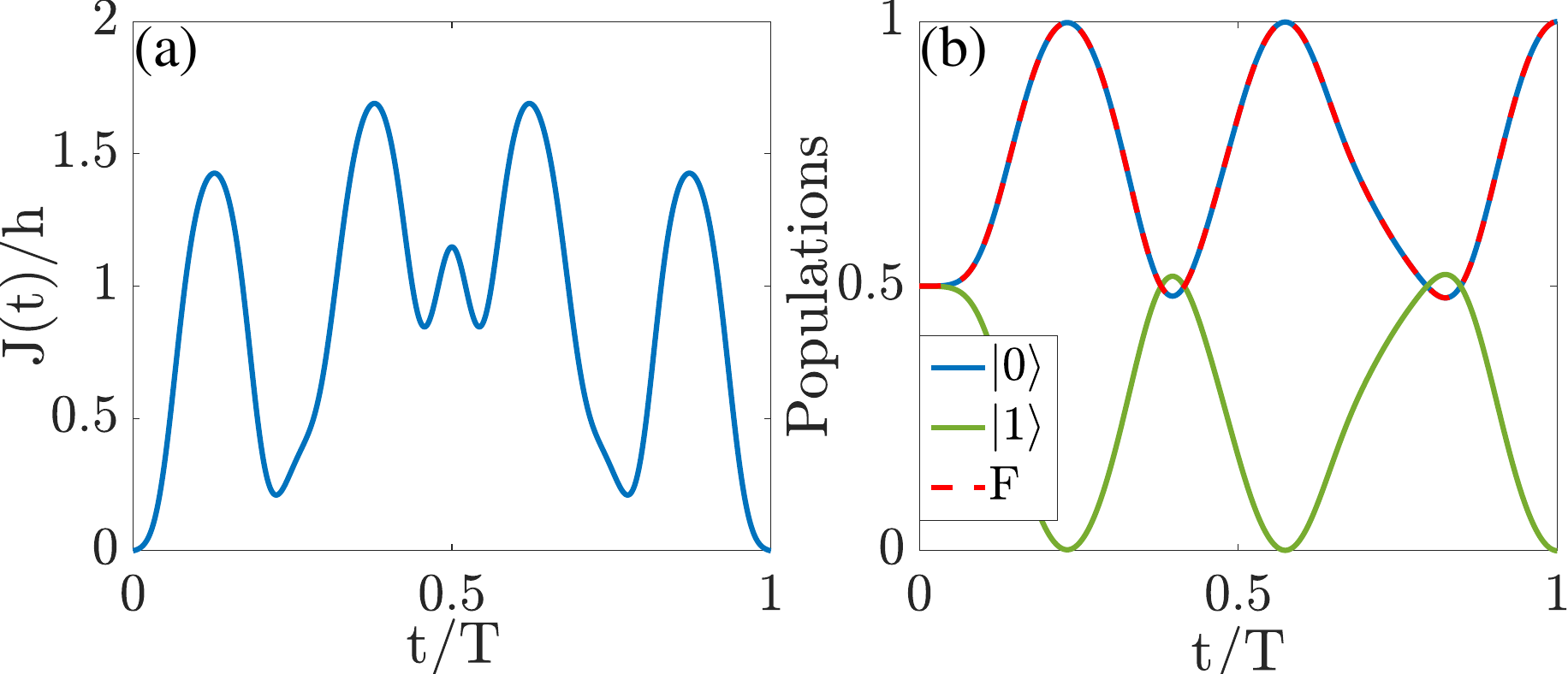}
  \caption{(a) The pulse shape of  $J(t)$ to implement Hadamard gate with a smooth curve. The parameters are $h/2= 2\pi$ MHz and $T=0.942$ $\mu s$. (b) Population of qubit states $\ket{0}$ and $\ket{1}$ and the corresponding gate fidelity in implementing the Hadamard gate.}
  \label{3populations_Hadamard}
\end{figure}

As demonstrations, we show how to construct Hadamard(H) gate and S gate, which are the generators of the group of 24 Clifford gate operations. To achieve smooth pulses, we set the value of the time-dependent $J(t)$ to be zero at the start and end points for both gates.
For H gate, the $\pi$ rotation around the axis $x+z$, we choose the boundary condition of $\zeta(0)=\zeta(t)=3\pi/8$, numerically solve equations to ensure $\xi_{\pm}=\pi/2$. As a specific example, we choose the trigonometric series of $\zeta(t)$ in Eq. (\ref{sin}) as
\begin{eqnarray}
\zeta(t)=A_{0}+A_{2}\cdot \sin^{2}\left(a_{2}\pi\frac{t}{T}\right)+A_{3}\cdot \sin^{3}\left(a_{3}\pi\frac{t}{T}\right),
\end{eqnarray}
with parameters $\{A_{0}, A_{1}, A_{2}, a_{2}, A_{3}, a_{3}\}=\{3\pi/8, 0, -0.22, 4, 0.18, 1\}$. Under these settings, a H gate is implemented. The pulse shape of $J(t)/h$, the corresponding state population and the gate-fidelity dynamics of this case are shown in Fig.\ref{3populations_Hadamard}.

Besides, to realize S gate, or any other phase gates, the decomposition in ST qubit system is
\begin{eqnarray}
R(\sigma_{z},\xi)=HR(\sigma_{x},\xi)H,
\end{eqnarray}
where $R(\sigma_{z/x},\xi)$ describes a rotation around axis $\sigma_{z}/\sigma_{x}$ in the Bloch sphere. To obtain this phase gate, an arbitrary rotation around the axis $\sigma_{x}$ is needed. According to Eq. (\ref{sin}), it could be realized by setting $\zeta(t)$, according to Eq. (\ref{sin}) as,
\begin{eqnarray}
\zeta(t)=A_{0}+A_{3}\cdot \sin^{3}\left(a_{3}\pi\frac{t}{T}\right),
\end{eqnarray}
The parameter $\xi$ could be solved numerically to satisfy $\xi=\pi/4$ or $\xi=\pi/8$ to implement S or T gate, respectively. As an example, we could set the parameters $\{A_{0},A_{1},A_{2},A_{3},a_{3}\}=\{\pi/4,0,0,0.24,1\}$ to perform $R(\sigma_{x},\xi=\pi/4)$. By combining this gate with two H gates, we obtain an S gate. Since the composition of H and S gates can generate all 24 Clifford gates, any randomized benchmarking (RB) process can be realized using only smooth pulse gates.

\begin{figure}[tbp]
  \centering
\includegraphics[width=\linewidth]{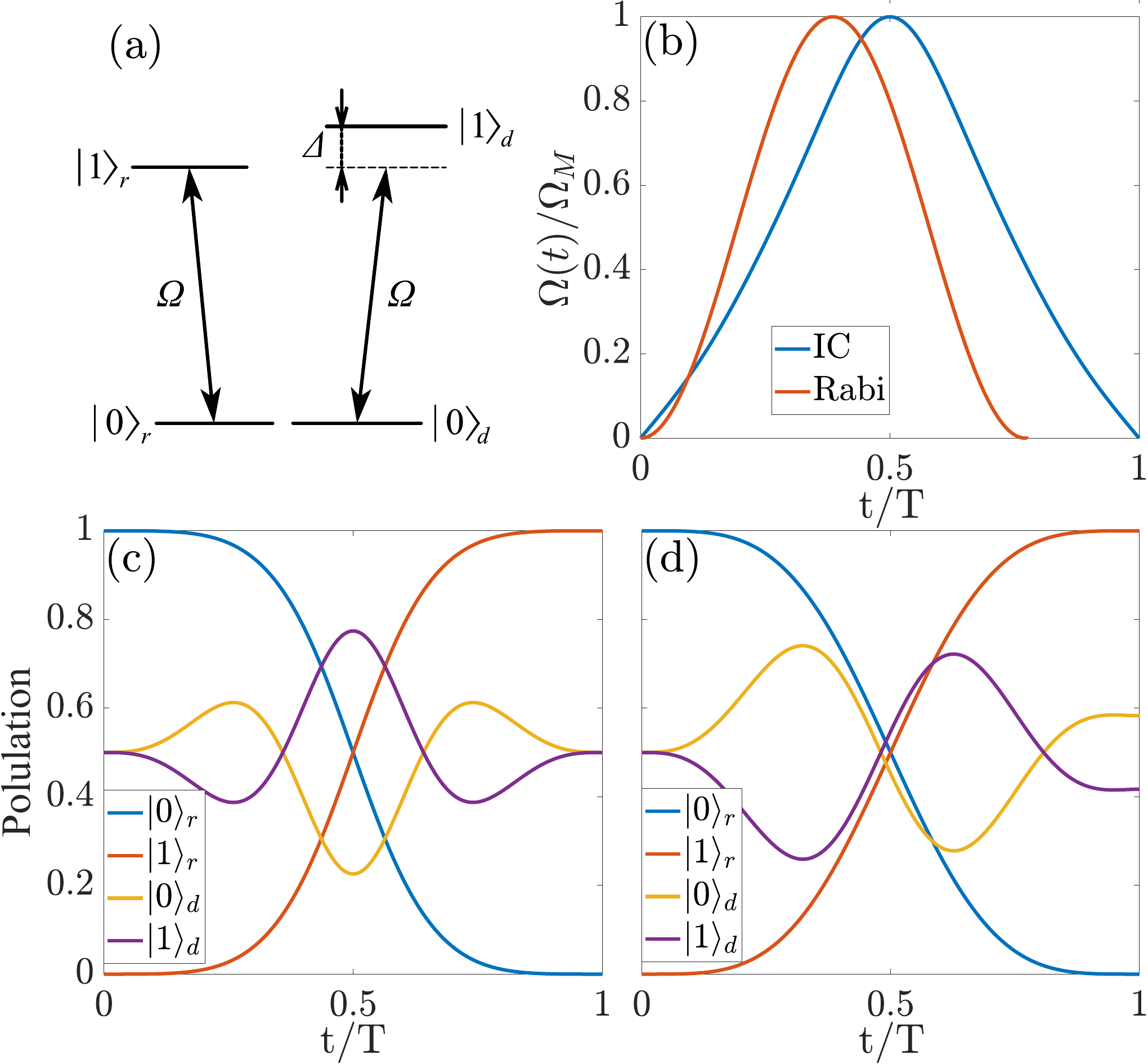}
\caption{Illustration of the example of individual control in a multilevel quantum system. 
(a) The energy level structure shows the scenario where the control pulse will also address the two-level subsystem in a detuned way, when two transitions are closely spaced in frequency. The resonant driving of the target subspace affects nearby transitions, leading to leakage in the target quantum dynamics.
(b) This panel compares the pulse shapes for individual control (IC) versus a general Rabi pulse in the form of the sine function, i.e., $\Omega_M \sin{(\pi t/T})$, where two pulses have same maximum amplitude $\Omega_M$, while T represents the evolution time for the individual control pulse.
(c) Plots of the populations of both the resonant and detuned subspaces for individual control, with parameters $\Delta=2\pi$ MHz, $T=0.95$ $\mu s$ and $\zeta(t)= \pi/8-0.29\cdot \sin^{3}{(\pi\cdot t/T)}$. Using an analytical-assisted solution, an accurate NOT operation is achieved without leakage to the nearby transition at the final time.
(d) The plots of the populations of both resonant and detuning subspaces for the general Rabi pulse $\Omega_M \sin{(\pi t/T})$, with the same parameters as in (c). While a NOT gate is obtained, the leakage of the control pulse results in unwanted evolution in the nearby transition.}
\label{fig3}
\end{figure}

\subsection{Individual controlling}

In many quantum control systems, there are numerous nearby transitions with comparable frequencies. Consequently, when implementing a desired manipulation on a target transition, there will inevitably be leakage to nearby transitions with different detuning.
This is also the case for some qubit systems, such as NV-center systems \cite{nv} trapped ions \cite{ions}, and molecular systems \cite{molecular}.
We here show an example of individual control over two nearby transitions, by using our analytical-assisted solution, and the corresponding energy level structure is presented in Fig. \ref{fig3}(a). This driving Hamiltonian on both transitions could be written as
\begin{subequations} 
\begin{align}
\label{Hamiltonian_CNot_2}
&H_{r}=\begin{pmatrix}
0 & \Omega(t) e^{-i\varphi}\\
\Omega(t) e^{i\varphi} & 0
\end{pmatrix},\\
&H_{d}=\begin{pmatrix}
\Delta & \Omega(t)e^{-i\varphi}\\
\Omega(t)e^{i\varphi} & -\Delta 
\end{pmatrix}.
\end{align}
\end{subequations}

Furthermore, another application of our solution is the ability to achieve precise control over both the target subspace and the nearby subspace. This allows for using one pulse to control two subspaces individually.
Specifically, In the target subspace, a resonant Rabi process is implemented by designing the Rabi rate $\Omega(t)$ and the phase $\varphi$. The integral area of $\int_0^T \Omega(t) dt$ is then calculated to realize the desired quantum gates, where $T$ labels the operation duration. This Rabi process with different phase $\varphi$ could obtain universal control over the target resonant subspace. Notice the quantum gates in this resonant subspace do not restrict the pulse shape but only set the pulse integrals. Meanwhile, for the nearby detuned subspace, we can design the pulse shape to match the analytical-assisted solution and achieve universal quantum control over it, as long as we do not change the value of the pulse integral. As the result, an arbitrary individual control over two transitions can be obtained.

Now, we show the example to obtain individual controlling. 
The choice is $\zeta(0)=\zeta(T)=\pi/4$ with free $\varphi(t)$, it can implement arbitrary phase gates, as shown in Fig. \ref{fig1}.
The corresponding evolution operator is 
\begin{subequations}
\begin{align}
\label{Operator_CNot_2}
&U_{r}=\begin{pmatrix}
\cos{[\int \Omega dt]} & -i\sin{[\int \Omega dt]} e^{-i\varphi}\\
-i\sin{[\int \Omega dt]} e^{i\varphi} & \cos{[\int \Omega dt]}\\
\end{pmatrix},\\
&U_{d}=\begin{pmatrix}
e^{-i\xi}      &0 \\
0     &e^{i\xi} 
\end{pmatrix}.
\end{align}
\end{subequations}
Since the phase  $\xi$ depends on $\zeta(t)$, we can design $\zeta(t)$  to implement a desired phase gate. Such as, an identity gate can be obtained when $\xi=\pi$. We show numeral simulations of the case, which implementing an individual control consisted of a NOT gate in resonant subspace and a phase gate in the detuned subspace, described in Fig. \ref{fig3}. Moreover, for the situation $\varphi=0$, another choice of $\zeta(t)$ can be used, it is $\zeta(0)=\zeta(T)=\pi/6$. It implements a rotation gate around the axis $\sigma_{x}$ in the resonant subspace and a gate determined by the value of $\xi$ in the nearby detuned subspace. The corresponding evolution operator is  
\begin{subequations} 
\begin{align}
\label{Operator_CNot_3}
&U_{r}=\begin{pmatrix}
\cos{[\int \Omega dt]} & -i\sin{[\int \Omega dt]} \\
-i\sin{[\int \Omega dt]}  & \cos{[\int \Omega dt]}\\
\end{pmatrix},\\
&U_{d}=\begin{pmatrix}
\cos{\xi}+i\frac{\sqrt{3}}{2}\sin{\xi}       &\frac{i}{2}\sin{\xi}\\
\frac{i}{2}\sin{\xi}     &\cos{\xi}-i\frac{\sqrt{3}}{2}\sin{\xi}
\end{pmatrix}.
\end{align}
\end{subequations}
If we design $\zeta(t)$ to satisfy  $\xi(T)=\pi$ and ensure $\int_0^T \Omega(t) dt=\pi$, we can realize an individual control gate consisting of a NOT gate in the resonant subspace and an identity gate in the detuned subspace. As known, a direct square pulse implementing the same individual control gate requires $\Omega=\Delta/\sqrt{3}$, which can be reproduced by this analytical solution case when setting $\zeta(t)=\pi/6$ for the whole time.

\section{Conclusion} 
In conclusion, we present an analytical-assisted solution of time-depended Schr\"{o}dinger equation for two level quantum system under driving, with few limitations. We show the details of the analytical progress and present some concrete examples to demonstrate its application in quantum control, i.e., deriving smooth pulse  for the gate operation in ST qubit systems and individual control over two transitions with nearby frequencies.

Further exploration in the field maybe the following. First, this solution could also take the gate robustness into account by choosing different free parameters. 
Second, it is important to extend the study to higher-level systems, such as three-level systems, which can be used to model superconducting transmon qubits. This involves incorporating new operators specific to three-level systems and reproducing the theoretical derivations.
Finally, the results presented here can also be extended to open quantum systems. By employing the Lindblad equation to simulate the density matrix, we ensure that the dynamics we obtain remain valid.

\acknowledgments
This work was supported by the National Natural Science Foundation of China (Grant No. 12275090) 
and the Innovation Program for Quantum Science and Technology (Grant No. 2021ZD0302303).

\end{document}